\documentclass[runningheads]{llncs}
\usepackage[T1]{fontenc}
\usepackage{amsmath}
\newcommand\asteriskfill{\leavevmode\xleaders\hbox{$\ast\ $}\hfill\kern0pt}
\usepackage{graphicx}
\usepackage{amsfonts}
\usepackage{makecell}
\usepackage{relsize}
\usepackage{xcolor}

\begin{document}
\title{LOTUS: Learning to Optimize Task-based US representations}

\author{Yordanka Velikova\inst{1} \and
 Mohammad Farid Azampour\inst{1,2} \and
 Walter Simson\inst{3} \and
 Vanessa Gonzalez Duque\inst{1,4} \and
 Nassir Navab\inst{1,5}
}

\authorrunning{Y. Velikova et al.}
\institute{Computer Aided Medical Procedures, Technical University of Munich, Germany \\ 
 \and
 Department of Electrical Engineering, Sharif University of Technology, Tehran, Iran \\ 
  \and
 Department of Radiology, Stanford University School of Medicine, Stanford, USA \\ 
 \and
 LS2N laboratory at Ecole Centrale Nantes, UMR CNRS 6004, Nantes, France\\
 \and
 Computer Aided Medical Procedures, John Hopkins University, Baltimore, USA \\
 }
\titlerunning\space{LOTUS: Learning to Optimize Task-based US representations}
\maketitle              
\begin{abstract}
Anatomical segmentation of organs in ultrasound images is essential to many clinical applications, particularly for diagnosis and monitoring. 
Existing deep neural networks require a large amount of labeled data for training in order to achieve clinically acceptable performance. Yet, in ultrasound, due to characteristic properties such as speckle and clutter, it is challenging to obtain accurate segmentation boundaries, and 
precise pixel-wise labeling of images is highly dependent on the expertise of physicians. In contrast, CT scans have higher resolution and improved contrast, easing organ identification.
In this paper, we propose a novel approach for learning to optimize task-based ultrasound image representations. Given annotated CT segmentation maps as a simulation medium, we model acoustic propagation through tissue via ray-casting to generate ultrasound training data. 
Our ultrasound simulator is fully differentiable and learns to optimize the parameters for generating physics-based ultrasound images guided by the downstream segmentation task. In addition, we train an image adaptation network between real and simulated images to achieve simultaneous image synthesis and automatic segmentation on US images in an end-to-end training setting. The proposed method is evaluated on aorta and vessel segmentation tasks and shows promising quantitative results. Furthermore, we also conduct qualitative results of optimized image representations on other organs.

\end{abstract}

\section{Introduction}

Ultrasound (US) imaging is a widely used modality in medical diagnosis for screening and follow-up examinations. Hence, precise segmentation of the target organs is crucial for diagnosing or tracking disease progression.
Recently, the application of deep learning for ultrasound image segmentation has emerged as a powerful tool. However, accurate segmentation of US images remains a challenging task due to the complexity of the modality, as it has limited resolution and often contains clutter, shadowing and reverberation artefacts.
This leads to a general lack of annotated data, and additionally, due to varying operator skills, there is high heterogeneity of ground truth data labels, which is the primary factor hampering solid segmentation performance~\cite{kronke2022tracked}.

On the other hand, large pixel-level labeled CT datasets are freely available online. 
Thus to overcome the lack of ground truth ultrasound data, researchers have utilized ultrasound simulators to generate large sets of ultrasound-like images from CT label maps and use them for training~\cite{ComparisonUSsim}.
Simulated ultrasound data automatically provides a labeled pair of the tissue distribution and the resulting b-mode image and can be augmented with rotational, brightness, contrast, probe, and scanner variations.

Generally, ultrasound simulators can be categorized into two types based on their modeling techniques: finite difference models of the wave equation, modeling the mechanical propagation of sound waves through tissues, and simulating ray casting through tissue maps represented by ultrasound tissue properties~\cite{jensen1997new,burger2012real,salehi2015patient}.
Although the former can model higher-order non-linear effects, producing realistic images, generating a single image can take hours. The latter, on the other hand, is much faster and can be integrated into other systems~\cite{hyun2019beamforming,brickson2021reverberation}. 
While leveraging automatically generated ultrasound simulations with corresponding labels for training has benefits, models trained on simulations fail when applied directly to real, as they cannot perfectly simulate ultrasound images without distinguishable differences from real ones.

Thus, one main challenge when working with simulated data is reducing the domain shift between simulated and real data.
In a supervised sense, many works have investigated the realistic parametrization of ultrasound simulators to reduce the domain shift between simulated and real data ultrasound data~\cite{simson2023investigating} and augmentation of ultrasound b-modes~\cite{tirindelli2021rethinking}.
Recent domain adaptation models~\cite{CycleGAN2017,park_contrastive_2020} employing generative adversarial networks have shown promise in improving image synthesis in an unsupervised manner. Moreover, recent works show their application in combination with segmentation or registration tasks between X-ray and CT or MRI scans~\cite{XrayImg2img,Dou2018UnsupervisedCD}. 
Further works show their application in ultrasound by closing the real-simulation gap via translation from simulated images to "realistic" ones that match the target domain, thereby enabling the application of trained segmentation networks on real images~\cite{Tomar2021ContentPreservingUT,Vitale2019ImprovingRI}.

However, those methods require separate training for each part of the architecture, limiting the models' flexibility. Notably,~\cite{CACTUSS} proposes using an intermediate representation image with common properties between CT and US for the task of aorta segmentation. However, the intermediate image is not formulated differentiably but is statically calibrated, and the whole pipeline is not trained end-to-end.

\subsubsection{Contributions}
In this paper, we propose a novel approach for learning to optimize task-based ultrasound image representations. 
During training, we render an intermediate US image representation from segmented public CT scans and use it as input to a segmentation network. Our ultrasound renderer is fully differentiable and learns to optimize the parameters necessary for physics-based ultrasound simulation, guided by the downstream segmentation task. 
At the same time, we train an image style transfer network between real and simulated data to achieve simultaneous image synthesis as well as automatic segmentation on US images in an end-to-end training setting. In addition, no labels are required for the real ultrasound images, which are also unpaired with the simulated ultrasound images.
We evaluate our method on aorta and vessel segmentation. Our quantitative and qualitative results demonstrate that our method learns the optimal image for the task of interest.
The source code for our method is publicly available\footnotemark[1]\footnotetext[1]{\url{https://github.com/danivelikova/lotus}}.

\begin{figure}[t]
\includegraphics[width=\textwidth]{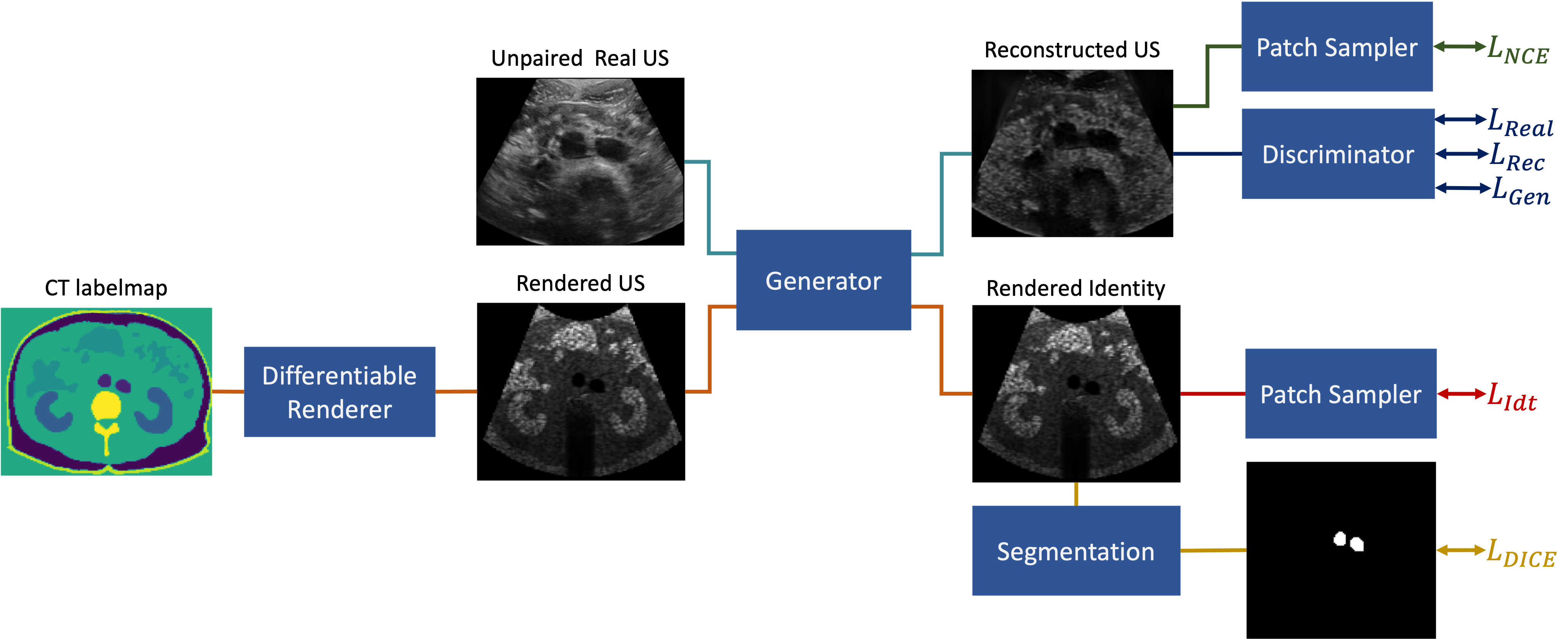}
\caption{Overview of the proposed framework. During training, we render online US simulation images from CT label maps and use them as input to a segmentation network. Our ultrasound renderer is fully differentiable and learns to optimize the parameters based on the downstream segmentation task.
At the same time, we train an unpaired and unsupervised image style transfer network between real and rendered images to achieve simultaneous image synthesis as well as automatic segmentation on US images in an end-to-end training setting.} \label{fig:overview}
\end{figure}

\section{Methodology}
\subsection{Differentiable Ultrasound Renderer}
\begin{figure}[t]
\includegraphics[width=\textwidth]{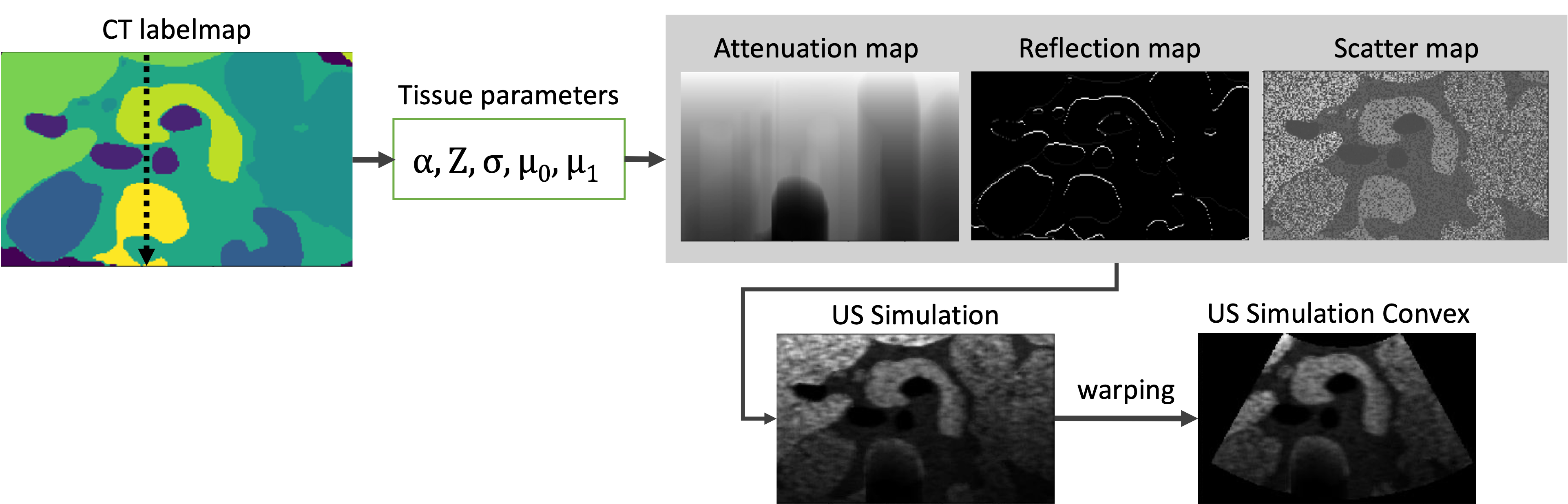}
\caption{Overview of the Differentiable Ultrasound Renderer.} \label{fig:us_renderer}
\end{figure}
Building on the mathematical foundations of ray tracing and ultrasound echo generation proposed by \cite{burger2012real}, we adopt those equations and modify them to be differentiable while still accurately depicting the physics behind generating US B-mode images.
Input to the renderer is a 2D label map with tissue labels. Each tissue label has assigned five parameters with default values\footnotemark[2]\footnotetext[2]{https://github.com/Blito/burgercpp/blob/master/examples/ircad11/liver.scene} which describe ultrasound-specific tissue characteristics and control the whole rendering generation - attenuation coefficient $\alpha$, acoustic impedance \textit{Z}, as well as three parameters that define the speckle distribution - $\mu_0$, $\mu_1$, $\sigma_0$. For each 2D label map, we use these parameters to define three sub-maps: attenuation, reflection, and scatter maps. 
We generate those maps by modeling ultrasound waves as rays starting from the transducer, which is the top of the label map, and propagating through media using physical laws. 
Ray casting is simulated by defining a function $E_i(d)$ for each scanline \textit{i} at a distance \textit{d} from the transducer, which describes the recorded ultrasound echo signal as:
\begin{equation}
E_i(d) = R_i(d) + B_i(d)
\end{equation}
Where $R_i(d)$ is the energy reflected from the interfaces between two tissues as the beam passes through them and $B_i(d)$ represents the energy backscattered from the scattering points along the scanline.
The reflection of the ray is described as:
\begin{equation}
R_i(d) = \left|I_i(d) * Z_i(d) \right| * P(d) \otimes G(d)
\end{equation}
where $I_i(d)$ is the remaining energy of the ray, which gets attenuated during tissue traversal.
We model $I_i(d)$ by approximating the Beer-Lambert Law as: $I_i(d) = e^{-\alpha d}$, where $\alpha$ is the attenuation coefficient of the medium and $d$ the distance travelled.
To construct the final 2D attenuation map, we calculate, for each ray, the cumulative product of the attenuation as it traverses through various tissues, thereby modeling how the signal's strength diminishes.
The reflection coefficient $Z = (Z_2 - Z_1)^2/(Z_2 + Z_1)^2$, is computed from the acoustic impedances of two adjacent tissues: $Z_1$ and $Z_2$. 
The $P(d)$ is the Point Spread Function (PSF) along the ray, and $G(d)$ is a boundary map, where 1 is assigned for points on the boundary of the surface and 0 otherwise. For simplicity, we model the PSF as a two-dimensional normalized Gaussian. 
The amount of the reflected signal, denoted by $\phi_r$, equals the result of multiplying the reflection coefficient by the boundary condition. To build our final 2D reflection map, for each ray, we compute the cumulative product of the residual signal, defined as $1 - \phi_r$. The output represents the fraction of the signal that propagates forward.

In additionally to the reflection term, a backscattered energy term $B_i(d)$ in the returning echo is calculated:
\begin{equation}
B_i(d) = I_i(d) * P(d) \otimes \widetilde{T}(x,y)
\end{equation}
the residual ultrasound wave energy $I_i(d)$ is multiplied with the PSF $P(d)$, which has been convolved with a texture $\widetilde{T}$ of random scatterers for each $(x,y)$, where:
\begin{align}
\widetilde{T}(x,y) &=
\begin{cases}
S(x,y) & \text{if } T_1(x,y) \leq \mu_1(x,y) \\
0 & \text{otherwise} \\
\end{cases} \\
S(x,y) &= T_0(x,y) * \sigma_0 + \mu_0
\end{align}
This texture is constructed using two random textures $T_0(x,y)$ and $T_1(x,y)$ with Gaussian normalized distributions and the parameters $\mu_ 0$, $\mu_1$, and $\sigma_0$, which represent the brightness, density and standard deviation of scatterers respectively.
To make the function fully differentiable, we replace the conditional operation \(T_1 \leq \mu_1\) with a differentiable approximation: 
\begin{align}
\widetilde{T}(x,y) &= \sigma(\beta \cdot (\mu_1(x,y) - T_1(x,y))) \cdot S(x,y)
\end{align}
where \(\sigma(z) = \frac{1}{1 + e^{-z}}\) is the sigmoid function and \(\beta\) is a scaling factor that adjusts its steepness.
The resulting function is fully differentiable as the sigmoid function smoothly approximates the step function and all operations involved are differentiable. 
Additionally, we apply temporal gain compensation (TGC) to enhance tissues deeper in the image. 
The final rendered ultrasound image is constructed from the three sub-maps (see Fig.~\ref{fig:us_renderer}) and additionally warped to produce the desired fan shape.
At the beginning of the training, we set the default tissue-specific values, which during the training, get changed, guided from the downstream task, and generate optimal US simulation.

\subsection{End-to-End Learning }
The proposed method's architecture is shown in Figure~\ref{fig:overview}. During training, our method follows two main paths: Real $\rightarrow$ Reconstructed US and CT label map $\rightarrow$ Segmentation. We explain the meaning of these paths in the order shown in the figure.

\textbf{Real $\rightarrow$ Reconstructed US}. 
Since there is an appearance gap between real and our rendered ultrasound images we incorporate an unpaired and unsupervised image-to-image translation network, CUT~\cite{park_contrastive_2020}, which uses a contrastive learning scheme.
Given a source image, the Generator learns a function $G: \mathcal{X} \mapsto \mathcal{Y}$ that translates the corresponding image into the target's appearance. We have two domains of unpaired instances: real US images as the source $\mathcal{X}$ and rendered US as the target $\mathcal{Y}$. The generator's encoder $G_{enc}$ extracts relevant content characteristics, while the decoder $G_{dec}$ learns to create the desired goal appearance.
The Generator network employs an adversarial loss:
\begin{equation}
    \mathcal{L}_{GAN} = \mathbb{E}_y \log D(y) + \mathbb{E}_x \log(1 - D(G(x)))
\end{equation}
where the generated images $G(x)$ resemble images from domain $\mathcal{Y}$, and $D(.)$ differentiates between translated and real images $y$.
However, the adversarial loss alone does not ensure that the translated image will preserve the structure of the anatomy.
An additional contrastive loss must be imposed, which
maximizes mutual information across corresponding image patches from the source and the output image.
We use the Patch Sampler from CUT to extract image patches and calculate the contrastive NCE ($\mathcal{L}_{NCE}$) loss~\cite{park_contrastive_2020}.
The final loss is defined as:
\begin{equation}
    \mathcal{L}_{CUT}(X,Y) = \mathcal{L}_{GAN}(X,Y) + \mathcal{L}_{NCE}(X, G(X)) + \mathcal{L}_{NCE}(Y, G(Y))
\end{equation}
where, the $\mathcal{L}_{NCE}$ is calculated on two pairs, a sample from the source domain ($x$) paired with the generated output $G(x)$ and a sample from the target domain ($y$) paired with the $G(y)$ which we denote as the identity image.
The loss over the second pair serves as an identity loss and prevents the generator from making unnecessary changes to the image.

\textbf{CT labelmap $\rightarrow$ Segmentation}: The segmentation network forward pass has a nested structure. First, we obtain a 2D slice from the CT label map and pass it to the differentiable ultrasound renderer. The resulting rendered US is passed through the frozen Generator network, and the identity image output of the Generator is used as an input to the segmentation network to ensure the same distribution as the target domain. We update both the segmentation network and the Renderer using dice loss. The label for computing the dice loss comes directly from the input label map used for generating the rendered US.

\textbf{Stopping criterion}: Once the segmentation network validation loss converges, we employ a small subset of 10 labeled images from the real US domain as a stopping indicator for the entire training pipeline.

\section{Experimental Setup}
\label{sec:data} 

\textbf{CT dataset:} We use 12 CT volumes from a publicly available dataset Synapse\footnotemark[3]\footnotetext[3]{https://www.synapse.org/\#!Synapse:syn3193805/wiki/89480}~\cite{landman2015miccai} for training.
The data comes with labels for multiple organs. These labels were additionally augmented with labels of bones, fat, skin, and lungs using TotalSegmentor~\cite{wasserthal2022totalsegmentator} to complete the label maps.\\
\textbf{In-vivo images:} We acquired abdominal ultrasound sweeps from eleven volunteers of age $26\pm3$ (m:7/f:4). For each person, one sweep was acquired with a convex probe\footnotemark[4]\footnotetext[4]{cQuest Cicada US scanner, Cephasonics, Santa Clara, CA, US}. Per sweep, 50 frames were randomly sampled and used for training the CUT network. To compare against a supervised approach, additional images were annotated (500 for the aorta, 400 for vessels) from all volunteers to train 5-fold cross-validation. From each set of annotated images, 100 images were randomly sampled as test sets for both segmentation tasks.

\noindent \textbf{Training Details:}
We train the network with a learning rate of $10^{-5}$ for the segmentation network, $10^{-3}$ for the US Renderer, and $5^{-6}$ for the image adaptation network, with a batch size of 1, Adam optimizer and dice loss. We employ rotation, translation, and scaling augmentations on the CT label maps and split them randomly in an 80-20\% ratio for training and validation, respectively.
For the supervised approach, we trained the networks, for 120 epochs, with a learning rate of $10^{-3}$ and the Adam optimizer. 

\begin{figure}[h]
    \centering
\includegraphics[width=0.95\textwidth]{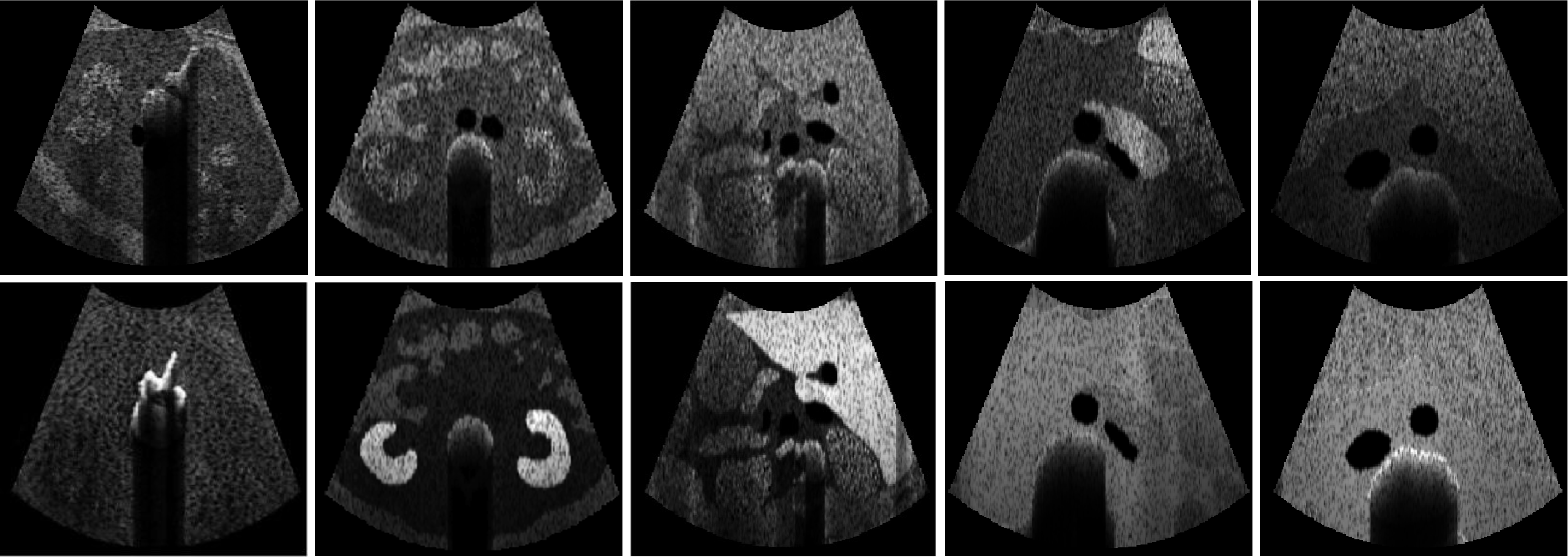}
\caption{Image representations of segmentation tasks for different target organs, learned during the optimization. Top row: rendered US with default parameters, bottom row: final optimized rendered US for each specific organ. From left to right: spine, kidney, liver, vessels, aorta only.} \label{fig:represnt}
\end{figure}

\noindent \textbf{Experiments}
We test the proposed framework quantitatively for two segmentation tasks: all vessels and the aorta only.
We evaluate the accuracy of the proposed method by comparing it to a supervised network. For this, we train a 5-fold cross-validation U-Net~\cite{Unet}, test on three hold-out subjects, and report the average DSC. We also compare to a fixed rendered image by freezing the US renderer instead of optimizing it. Additionally, we show qualitative results of the proposed method when the downstream tasks are changed Fig.~\ref{fig:represnt}.

\section{Results and Discussion}
In Table~\ref{tab:aorta_dice}, we compare the performance of LOTUS against a fully supervised approach and against a frozen renderer's parameters and report the DSC and Hausdorff distance (HD) in mm for aorta segmentation and DSC only for all vessels segmentation.
Our proposed method achieved the highest DSC score of $89.24\pm0.13$ and the lowest HD score of $2.52\pm1.18$ mm for aorta segmentation.
For the task of vessels segmentation it also achieved the best DSC of $90.9\pm0.06$.
\begin{table}[h]
\centering
    \caption{Comparison of DSC and Hausdorff distance for the task of aorta and vessels segmentation of our proposed method with supervised network and with frozen renderer.}
    \begin{tabular}{ l | p{60pt} | p{70pt} | p{70pt} }
      & \thead{Supervised}  & \thead{Frozen Renderer} & \thead{LOTUS}\\ 
     \hline
     DSC - Aorta & 80.65$\pm$2.35 & 84.67$\pm$0.14  & $\mathbf{89.24\pm0.13}$  \\
     \hline
     Hausdorff (mm) & 17.61$\pm$1.32 & 11.08$\pm$18.64 & $\mathbf{2.52\pm1.18}$  
\\ 
    \hline
     \hline DSC - Vessel & 83.56$\pm$4.16 & $89.05\pm0.09$ & $\mathbf{90.9\pm0.06}$ \\
    \end{tabular}
    \label{tab:aorta_dice}
\end{table}
Fig. \ref{fig:represnt} depicts the images obtained during the optimization of the proposed method for different target organs. The upper row shows the rendered US image with default parameters, and the bottom row displays the optimized image representations for the corresponding target organ learned during optimization. It can be observed that the spine, kidney and liver appear brighter, while for vessel and aorta segmentation, the vessels darken and the background becomes uniformly homogeneous. This highlights the ability of the proposed method to learn optimal representations for each downstream task.

The results presented in this work demonstrate the effectiveness of LOTUS for segmenting organs in ultrasound images. Our physics-based simulator generates synthetic training data, which 
is especially useful in scenarios where obtaining labeled data is time-consuming or costly. 
We believe that learning from transferred labels from CT contributes to a more accurate model since CT data is more accessible and labels are more refined.
Our quantitative results indicate that LOTUS can achieve accurate segmentation of aorta boundaries and other vessels. 
Furthermore, the end-to-end framework enables the differentiable US renderer and the unsupervised image translation to get optimized dynamically during the training.
Thus, the intermediate representation image is not static but changes during the training.
This illustrates the adaptivity of the proposed method to the downstream task, highlighting its prospective applicability across diverse applications and anatomies. 

Moreover, rather than directly using the rendered US image as an input to the segmentation network, we use the identity image output from the Generator. This yielded significant improvement in the segmentation result as it learns from a distribution consistent with the reconstructed US while looking similar to the rendered US. As a result, during inference stage, the distribution of the translated real US
is closer to the distribution the segmentation network was trained on, thereby improving the performance of the model.

One of the challenges when employing generative adversarial networks is that the loss is not an indicator of the best result.
We determine the optimal model by utilizing a small subset of labeled images after the convergence of the segmentation network, to ensure robustness during inference. Further stopping criteria can be studied to achieve higher automation of the pipeline.

Currently, our model incorporates the basic physics of ultrasound imaging without considering artifacts explicitly. Thus, exploring the robustness of the method against artifacts could yield valuable future improvements.

\section{Conclusion}
This paper presents a novel approach to learning task-based ultrasound image representations. LOTUS leverages CT labelmaps to simulate ultrasound data via differentiable ray-casting. The proposed ultrasound simulator is fully differentiable and learns to optimize the parameters for generating physics-based ultrasound images guided by the downstream segmentation task.
We also introduce an image adaptation network to achieve simultaneous image synthesis and automatic segmentation on US images in an end-to-end training setting without needing paired real and simulated images.
Our method is evaluated on aorta and vessel segmentation tasks and shows promising quantitative results. Furthermore, we demonstrate the potential of our approach for other organs through qualitative results of optimized image representations.
The ability to learn from unlabeled data and simulate the ultrasound modality has the potential for various clinical tasks beyond segmentation. We believe that our work has the potential to improve ultrasound imaging interpretation and learning.

\section*{Acknowledgements}
We would like to thank Magdalena Wysocki for the insightful discussions and Dr. Magdalini Paschali for helping with refining and improving the manuscript. The authors were partially supported by the grant NPRP-11S-1219-170106 from the Qatar National Research Fund (a member of the Qatar Foundation). The findings herein are however solely the responsibility of the authors.

%
%
\bibliographystyle{splncs04}
\bibliography{bibliography}
%

\appendix
\leavevmode\newline
%
\def\etal{~\textit{et.al.}}
%


\section*{Supplementary Material} \leavevmode\newline

\subsection*{Additional Qualitative Results}
\begin{figure}[h]
    \centering
\includegraphics[width=\textwidth]{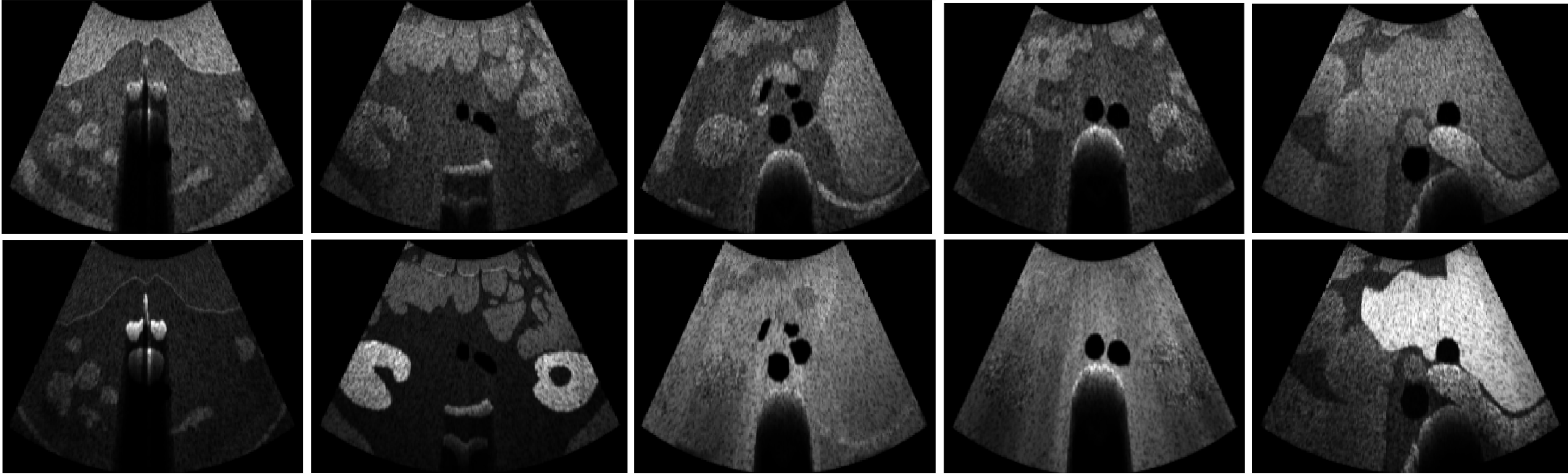}
\caption{Additional image representation results obtained after training for segmentation of different target organs. Top row: rendered US with default parameters, bottom row: final optimized US. The results are from five different segmentation tasks, from left to right: spine, kidney, vessels, aorta only, liver. It can be observed that the spine, kidney and liver get enhanced, while for vessels  and aorta only segmentation, the vessels get darker and the background more homogeneous. This demonstrates the ability of the proposed method to learn different representations based on the downstream task.
} \label{fig:augm}
\end{figure}

\begin{figure}[h]
    \centering
\includegraphics[width=\textwidth]{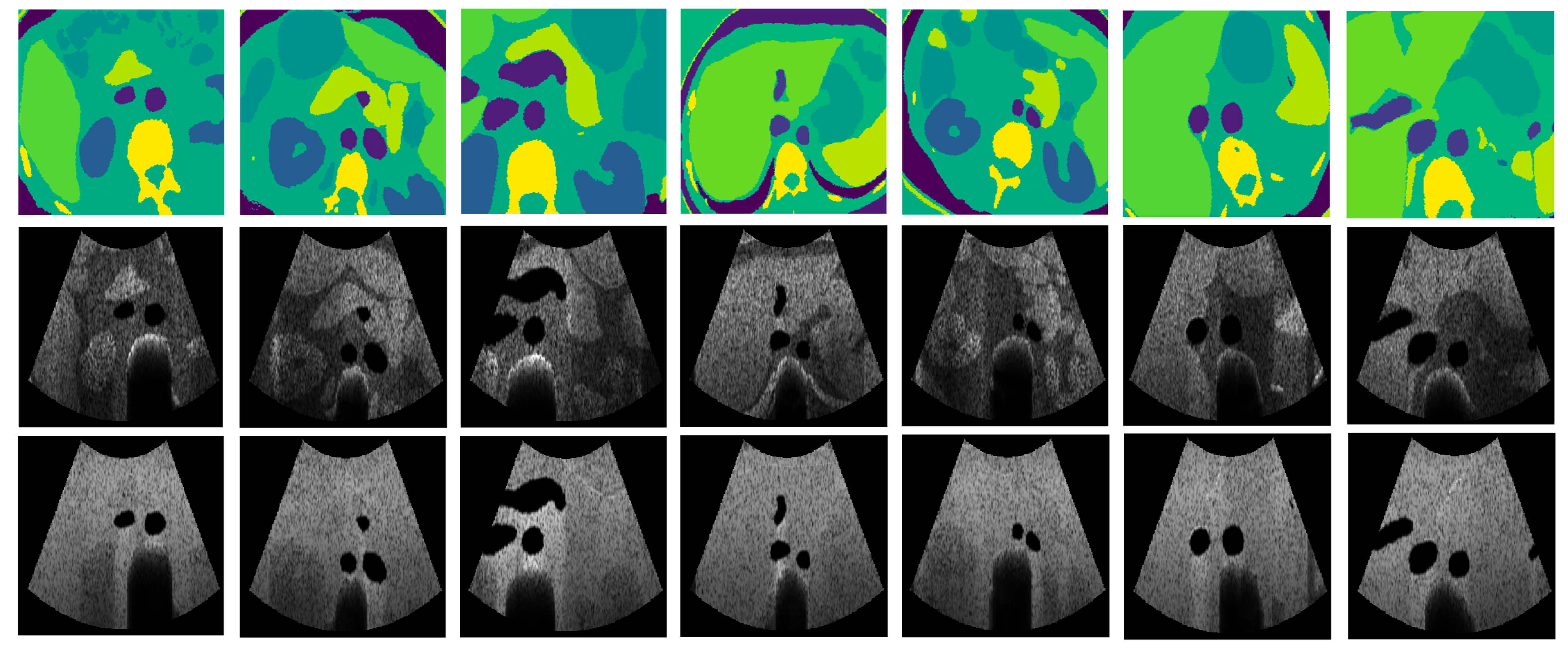}
\caption{Top: input labelmaps during training with different augmentations (rotation, translation, scaling). Middle: rendered US images with default parameters. Bottom: optimized US images after training for the task of vessels segmentation.} \label{fig:augm}
\end{figure}

\begin{figure}[h]
    \centering
\includegraphics[width=\textwidth]{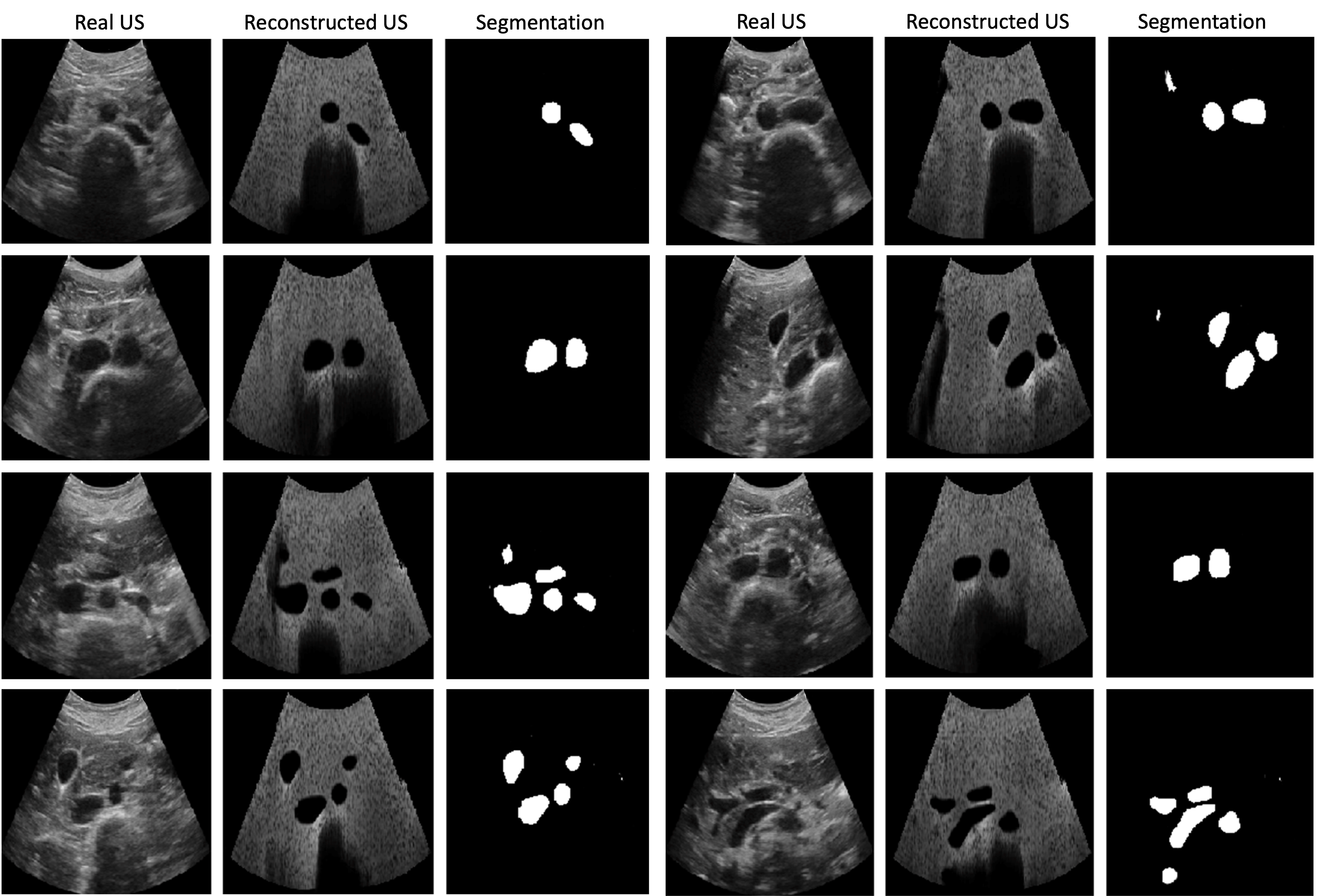}
\caption{Results during inference: real US images are first passed trough the image adaptation network and then through the segmentation network. The real US images are translated to the optimized US image appearance, which was learned during the training for the task of vessels segmentation.} \label{fig:real2reconstr}
\end{figure}

\end{document}